\documentclass[journal]{IEEEtran}

\usepackage{graphicx, dblfloatfix}
\usepackage{amsmath}
\usepackage{amssymb}
\usepackage{graphics}
\usepackage{cite}
\usepackage{amsfonts}
\usepackage{textcomp}
\usepackage{multicol}
\usepackage{multirow}
\usepackage{color}
\usepackage{fixltx2e}
\usepackage{epstopdf}
\usepackage{bm}
\usepackage{siunitx}
\usepackage{subfigure}

\begin{document}

\title{Cascode Cross-Coupled Stage High-Speed Dynamic Comparator in 65\,nm CMOS}

\author{Komala Krishna and Nandakumar Nambath
\thanks{Komala Krishna and Nandakumar Nambath are with the School of Electrical Sciences, Indian Institute of Technology Goa, Ponda - 403401, India (e-mail: krishna183422003@iitgoa.ac.in, npnandakumar@iitgoa.ac.in).}}

\maketitle

\begin{abstract}

  Dynamic comparators are the core of high-speed, high-resolution analog-to-digital converters (ADCs) used for communication applications. Most of the dynamic comparators attain high-speed operation only for sufficiently high input difference voltages. The comparators' performance degrades at small input difference voltages due to a limited pre-amplifier gain, which is undesirable for high-speed, high-resolution ADCs. To overcome this drawback, a cascode cross-coupled dynamic comparator is proposed. The comparator improves the differential gain of the pre-amplifier and reduces the common-mode voltage seen by the latch, which leads to a much faster regeneration at small input difference voltages. The proposed comparator is designed, simulated, and compared with the state-of-the-art techniques in a 65\,nm CMOS technology. The results show that the proposed comparator achieves a delay of 46.5\,ps at 1\,mV input difference, and a supply of 1.1\,V.
  
\end{abstract}


\IEEEpeerreviewmaketitle
\vspace{-0.3cm}
\section{Introduction}
\label{sec:intro}

  \IEEEPARstart{A}{nalog-to-digital converters} are widely used in various applications due to the increased demand for mixed-signal systems \cite{liu201712}. Comparator, an essential block in ADCs, plays a vital role in determining the speed and accuracy of the ADCs. The performance of an ADC relies on the robustness of the comparator \cite{hong20152}, especially for low noise, low-power, and high-speed operations. Dynamic comparators are preferred in low-power and high-speed designs due to their zero static power. They are classified as single-tail (ST) and double-tail (DT) topologies. Various ST circuits are reported to suffer from trade-offs between energy consumption (EC), offset, and speed \cite{devarajan201712}. ST topology also suffers from large kickback (KB) noise and requires a large voltage headroom since the input transistors are directly stacked with the cross-coupled pair. Due to these drawbacks, the DT configuration is preferred for the design of high-speed comparators \cite{li2020low}.
  
  Designing a high-speed comparator that can resolve small input difference voltages while holding on to the high-speed capability over a wide range of common-mode voltage is challenging \cite{khorami2018low}. The conventional DT comparator reported in \cite{schinkel2007double} has mitigated the drawbacks of the ST comparator. However, it fails to give valid outputs for small input difference voltages. This has a direct impact on the resolution of ADCs. Moreover, at higher common-mode levels, the performance of the conventional DT comparator degrades because the input pair enters the triode region without providing sufficient gain. A dynamic comparator resistant to common-mode variations with delayed operation of the latch is presented in \cite{gao2015high}. However, it requires a large area and suffers from increased KB noise. Further, the insufficient pre-amplifier gain makes it impractical to use in high precision ADCs. In the dynamic bias DT comparator presented in \cite{bindra20181} the pre-amplifier partially discharges the drains of the input transistor pair to reduce EC. However, the speed is compromised to attain energy efficiency. To improve the latch regeneration time, a transconductance-enhanced latch stage is presented in \cite{wang2019low}. It has the same drawback as the conventional DT comparator in its common-mode performance. Additionally, due to stacking in the latch stage, the delay increases swiftly for lower supply voltages.
  
  Our work targets to reduce the comparator delay by enhancing the pre-amplifier gain compared to other high-speed DT architectures reported. The performance improvement is achieved by including a cascode cross-coupled pair in the pre-amplifier stage. The circuit is designed and implemented in a 65\,nm CMOS technology with a 1.1\,V supply. The proposed technique offers better delay performance throughout the input voltage range, especially at smaller input differences. Also, the cascode cross-coupled pair alleviate the delay degradation at higher common-mode voltages. These advantages make the proposed comparator suitable for high-speed, high resolution ADCs.
  
  This paper is organized as follows. Section \ref{sec:conv} presents the conventional DT comparator and its operation. The proposed comparator and its delay analysis are provided in Section \ref{sec:prop}. Simulation results and discussions are presented in Section \ref{sec:results}, and the conclusion is given in Section \ref{sec:conclusion}.
\vspace{-0.4cm}
\section{Conventional Double Tail Dynamic Comparator}
\label{sec:conv}
\vspace{-0.15cm}

  The conventional DT dynamic comparator has an input stage and a latch stage that have separate tail transistors. Two independent tail currents enable us to optimize the design trading-off speed, offset, and EC. This topology has fewer transistors stacked \cite{zhang2020high}, making it suitable for low voltage applications. It also reduces the KB noise due to the isolation between the input transistors and the output nodes.

  In the conventional DT, at smaller input difference voltages ($\Delta V_{IN}$), the latch is unable to sense the differential voltage due to the limited differential gain of the pre-amplifier. The proposed comparator mitigates this drawback by lowering the common-mode voltage and improving the differential voltage at the pre-amplifier output. This helps the latch to regenerate faster even at smaller $\Delta V_{IN}$.
 
\section{Proposed Cascode Cross-Coupled Dynamic Comparator}
\label{sec:prop}

\begin{figure}[!t]
    \centering
    \includegraphics[scale=0.4]{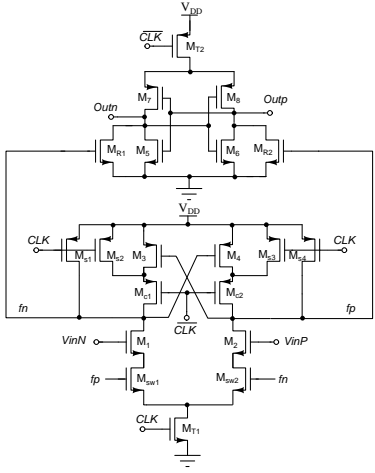}
    \caption{Schematic of the proposed double tail comparator with cascode cross-coupled pair to enhance pre-amplifier gain. The cascode cross-coupled pair made up of $\text{M}_\text{3}$, $\text{M}_\text{4}$, $\text{M}_\text{c1}$, and $\text{M}_\text{c2}$ improves the pre-amplifier performance.}
    \label{fig_prop}
  \end{figure}

  A PMOS cross-coupled pair is employed to increase the differential gain of the pre-amplifier in \cite{babayan2013analysis}. To enhance the performance further, the proposed topology, shown in Fig.~\ref{fig_prop}, introduces a cascode cross-coupled pair made up of $\text{M}_\text{3}$, $\text{M}_\text{4}$, $\text{M}_\text{c1}$, and $\text{M}_\text{c2}$. As a result, a higher difference voltage, $\Delta V_{fn,fp}$, at the pre-amplifier output nodes ($fn$, $fp$) is observed by the latch. This helps to reduce latch regeneration time and to resolve for smaller $\Delta V_{IN}$. 
  
\vspace{-0.2cm}

\subsection{Operation}
\label{subsec:operation}
  
    During the reset phase ($CLK=0$), the tail transistors $\text{M}_\text{T1}$ and $\text{M}_\text{T2}$ are off along with the cascode transistors  $\text{M}_\text{c1}$ and $\text{M}_\text{c2}$. The switching transistors $\text{M}_\text{s1}$ and $\text{M}_\text{s4}$ charge the $fn$ and $fp$ nodes to $\text{V}_{\text{DD}}$. Similarly, $\text{M}_\text{s2}$ and $\text{M}_\text{s3}$ charge the drain nodes of $\text{M}_\text{3}$ and $\text{M}_\text{4}$ to $\text{V}_{\text{DD}}$. Therefore, $\text{M}_\text{3}$ and $\text{M}_\text{4}$ are off. The transistors $\text{M}_\text{R1}$ and $\text{M}_\text{R2}$ ensure a proper start condition for the comparator. During the comparison phase ($CLK=\text{V}_\text{DD}$), $\text{M}_\text{T1}$ and $\text{M}_\text{T2}$ are on, and $\text{M}_\text{s1-s4}$  are off. In the beginning of this phase, the PMOS cascode cross-coupled pair is still inactive, and the transistors $\text{M}_\text{sw1}$ and $\text{M}_\text{sw2}$ are on. In this scenario, the operation of the pre-amplifier in the proposed circuit resembles that of its conventional counterpart.
    
  When $VinN>VinP$, the node $fn$ discharges faster and provides sufficient source to gate voltage for $\text{M}_\text{4}$ to turn on. 
  When $\text{M}_\text{4}$ turns on first, node $fp$ starts charging to $\text{V}_{\text{DD}}$ through $\text{M}_\text{c2}$. Simultaneously, $\text{V}_{\text{SG}}$ of $\text{M}_\text{3}$ falls below the threshold voltage, restricting the current through $\text{M}_\text{3}$ and $\text{M}_\text{c1}$ to the node $fn$. Hence, the node $fp$ charges to $\text{V}_{\text{DD}}$ and the node $fn$ discharges to 0\,V. The PMOS cascode cross-coupled structure increases the differential voltage $\Delta V_{fn,fp}$. Subsequently, the initial voltage difference sensed by the latch before regeneration, $\Delta V_O$, is improved as demonstrated below. NMOS transistor switches $\text{M}_\text{sw1}$ and $\text{M}_\text{sw2}$ take care of the static power dissipation in the pre-amplifier. They also contribute to the increased $\Delta V_{fn,fp}$ by manifesting another NMOS cascode cross-coupled pair with the input transistors $\text{M}_\text{1}$ and $\text{M}_\text{2}$. As a result, the latch regeneration time is decreased.
  
  To demonstrate the enhanced $\Delta V_{fn,fp}$, the procedure used in \cite{khorami2018low} is adopted. The delay analysis of the conventional DT comparator holds for the proposed comparator for most parts except for calculating the gain from input to the $fn$ and $fp$ nodes. 
\vspace{-0.3cm}
  
\subsection{Delay Analysis}
\label{subsec:delay}
 
  The delay of the proposed comparator is the sum of amplification time, $t_{amp}$, and the latch delay time, $t_{latch}$. The latch delay is given by \cite{khorami2018low} 
    \begin{equation}
    \begin{aligned}
      t_{latch} &=\tau_{inv}\times\ln{\frac{\text{V}_\text{DD}/2}{\Delta V_O}}+ \\ &\frac{C_L}{K_{5,7}(V_{DD}-V_{CML}-V_{Thp})^2}\times V_{Thn}
    \end{aligned}
  \end{equation}
  where $\tau_{inv}={C_L}/(G_{m5,6}+G_{m7,8}+G_{mR1,2})$, $V_{CML}$ is the common-mode voltage seen by the latch, and $K_{5,7}=0.5\times \mu_nC_{ox}(W/L)_{5,7}$. This equation indicates that a higher $\Delta V_{O}$, implying a higher $\Delta V_{fn,fp}$, and a lower $V_{CML}$ provide a smaller delay. The proposed comparator achieves both with the help of the cascode cross-coupled pair.
  
  To demonstrate the efficacy of the proposed comparator, the half circuit analysis can be used. Here, transistors $\text{M}_\text{3}$ and $\text{M}_\text{4}$ are modelled as current sources as they operate in the saturation region in the initial stages of the comparison phase. Transistors $\text{M}_\text{c1}$ and $\text{M}_\text{c2}$ are in the saturation region as their gate terminals are at ground potential.
  
  Applying KCL at nodes $fn$ and $fp$ gives the following expressions
  \begin{equation}
  \label{eqn:9}
    \begin{split}
      V_{fn}=(A_I g_{m3} V_{fp}-I_1)t_{amp}/C_P, \\
      V_{fp}=(A_I g_{m4} V_{fn}-I_2)t_{amp}/C_P
    \end{split}
  \end{equation}
  where $C_P$ is the parasitic capacitance at nodes $fn$ and $fp$, $g_{m3}=g_{m4}=g_{m}$, $I_1$ and $I_2$ are the drain currents of $\text{M}_\text{1}$ and $\text{M}_\text{2}$, respectively. $A_I$ is the current gain provided by the common gate stage formed by the cascode transistors $\text{M}_\text{c1}$ and $\text{M}_\text{c2}$, which is less than unity. Using the small signal analysis $A_I$ can be expressed as
   \begin{equation}\label{eqai}
      A_I= -\frac{[1/r_{c1,2}+(g_{mc1,2}+g_{mbc1,2})r_s]}{[1+1/r_{c1,2}+(g_{mc1,2}+g_{mbc1,2})r_s]}
  \end{equation}
  where $r_{c1,2}$ and $g_{mc1,2}$ are the channel resistance and transconductance of cascode transistors, respectively, and $r_s$ is the source resistance.
  
  By solving the linear equations (\ref{eqn:9}), we get
  \begin{equation}
    \begin{split}\label{eq_fnfp}
      V_{fn}=\frac{t_{amp}/C_P[g_mI_2(t_{amp}/C_P)+I_1]}{1-A_I g^2_m(t_{amp}/C_P)^2}, \\
      V_{fp}=\frac{t_{amp}/C_P[g_mI_1(t_{amp}/C_P)+I_2]}{1-A_I g^2_m(t_{amp}/C_P)^2}.
    \end{split}
  \end{equation}
  
  The common-mode voltage at the pre-amplifier output is given by $V_{CM,fn,fp}=(V_{fn}+V_{fp})/2$. By substituting (\ref{eq_fnfp}) in $V_{CM,fn,fp}$, we get
  \begin{equation}\label{eq11}
      V_{CM,fn,fp}= \frac{I(t/C_P)}{1-A_I g_m(t/C_P)}
  \end{equation}
  where $I = I_1 + I_2$.
  
  The differential voltage at the pre-amplifier output nodes is expressed as $\Delta V_{fn,fp} = V_{fn}-V_{fp}$ and is given by
  \begin{equation}\label{eq12}
    \Delta V_{fn,fp}=\frac{\Delta I(t/C_P)}{1+A_I g_m(t/C_P)}
  \end{equation}
  where $\Delta I = I_1 - I_2$.

  From (\ref{eq11}) and (\ref{eq12}), the factor $A_I (<1)$ obtained due to the cascode cross-coupled pair reduces the common-mode voltage and improves the differential voltage at $fn$ and $fp$ nodes. This helps the latch to regenerate faster than the other comparator architectures. Reducing the $g_m$ of $\text{M}_\text{3}$ and $\text{M}_\text{4}$ will also achieve the same result but at the cost of an increased offset voltage.
\vspace{-0.2cm}
\section{Results and Discussion}
\label{sec:results}
   \begin{figure}[!t]
    \centering
    \begin{tabular}{l}
    \includegraphics[width=0.4\textwidth]{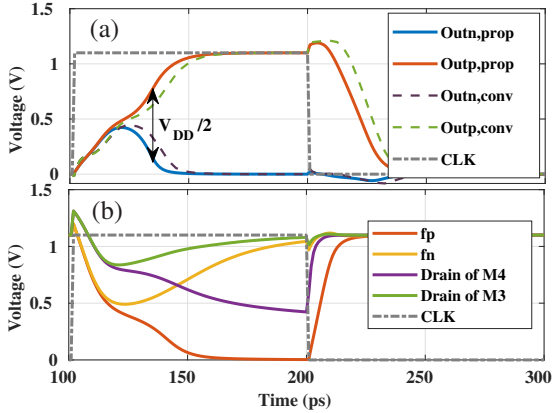} \\[-5.4cm] \hspace{1.1cm}(a) \\[2cm] \hspace{1cm} (b) \\[2.5cm]
    \end{tabular}
    \caption{Transient analysis results of the proposed comparator in comparison with the conventional one  when $\Delta V_{IN}=10$\,mV, $\text{V}_\text{CM}=0.77$\,V: (a) output nodes and (b) intermediate nodes.}
      \vspace{-0.4cm}

    \label{fig_trans}
  \end{figure}

  The performance metrics of analog circuits, in general, are technology-dependent. Dynamic comparators are no exception and are prone to both analog and digital non-idealities. Therefore, to make a fair comparison between the proposed and the reported dynamic comparator topologies, all the comparators are simulated in 65\,nm technology with a minimum channel length of 60\,nm, load capacitance of 2\,fF, $CLK$ of \SI{5}{\giga\hertz}, $\text{V}_\text{DD}$ of 1.1\,V, and $\text{V}_\text{CM}$ of 0.77\,V. It is also ensured that all the comparators are designed and optimized to obtain a similar offset standard deviation, $\sigma_{os}$. 

  Schematic level transient analysis results of the comparator is shown in Fig.~\ref{fig_trans}. The reference input, $V_P$, is fixed at 0.77\,V to attain the optimum performance \cite{babayan2013analysis}. The voltages at the output nodes, $Outp$ and $Outn$ and the intermediate nodes are shown at a $\Delta V_{IN}$ of 10\,mV. The delay is evaluated when the output node voltages attain a difference of $\text{V}_{\text{DD}}/2$. The delay of the proposed comparator is found to be 33.3\,ps whereas, it is 39.3\,ps for the conventional one. 
 
\begin{figure}[!t]
    \centering
    \includegraphics[scale=0.5]{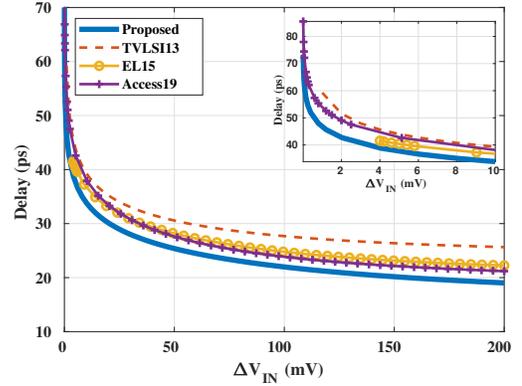}
    \caption{Delay variation with $\Delta V_{IN}$ of the proposed comparator. Inset shows the improved delay for small values of $\Delta V_{IN}$.}
      \vspace{-0.7cm}

    \label{fig_dv}
  \end{figure}

  Fig.~\ref{fig_dv} presents the delay variation with $\Delta V_{IN}$ at a $\text{V}_\text{CM}$ of 0.77\,V. The delay of the proposed comparator is significantly lower over the input difference voltage range. Especially, for lower $\Delta V_{IN}$ values, where the comparator presented in \cite{babayan2013analysis} and the conventional comparator fail to give valid outputs. It can be seen in the inset of Fig.~\ref{fig_dv}. This can be attributed to the cascode cross-coupled pre-amplifier, which provides sufficient $\Delta V_O$ for the latch to regenerate even at lower input difference voltages. From simulation results, it is observed that the proposed comparator is 25\,\% faster than the conventional comparator and 15\,\% faster than the \cite{babayan2013analysis} at a $\Delta V_{IN}$ of 1\,mV. This advantage allows us to incorporate this topology in high-speed, high-resolution ADCs provided proper care is taken in the design to reduce the input-referred noise. 
  
\begin{figure}[!t]
    \centering
        \subfigure[]{\includegraphics[width=0.24\textwidth]{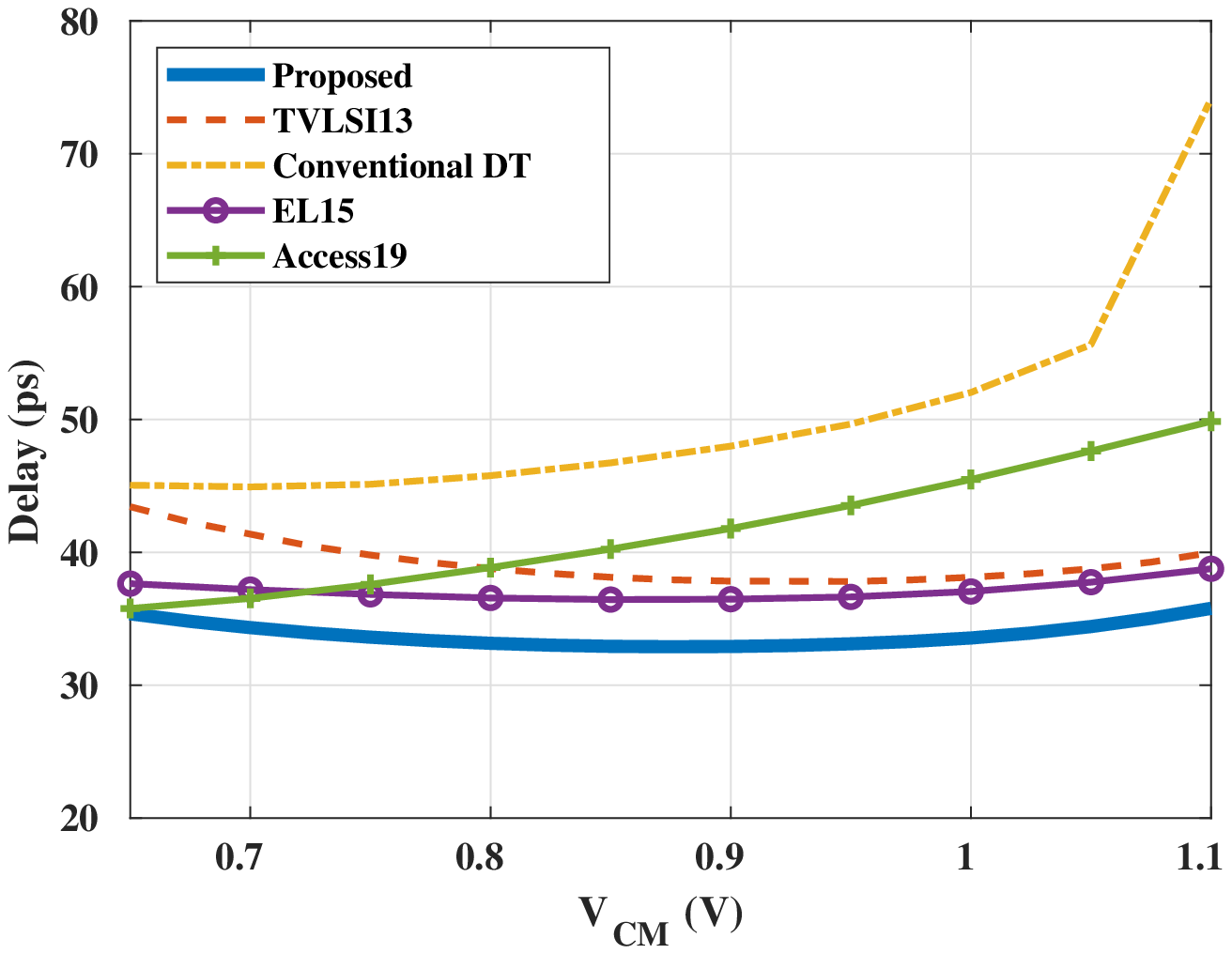}} 
    \label{lay:vid}
    \subfigure[]{\includegraphics[width=0.24\textwidth]{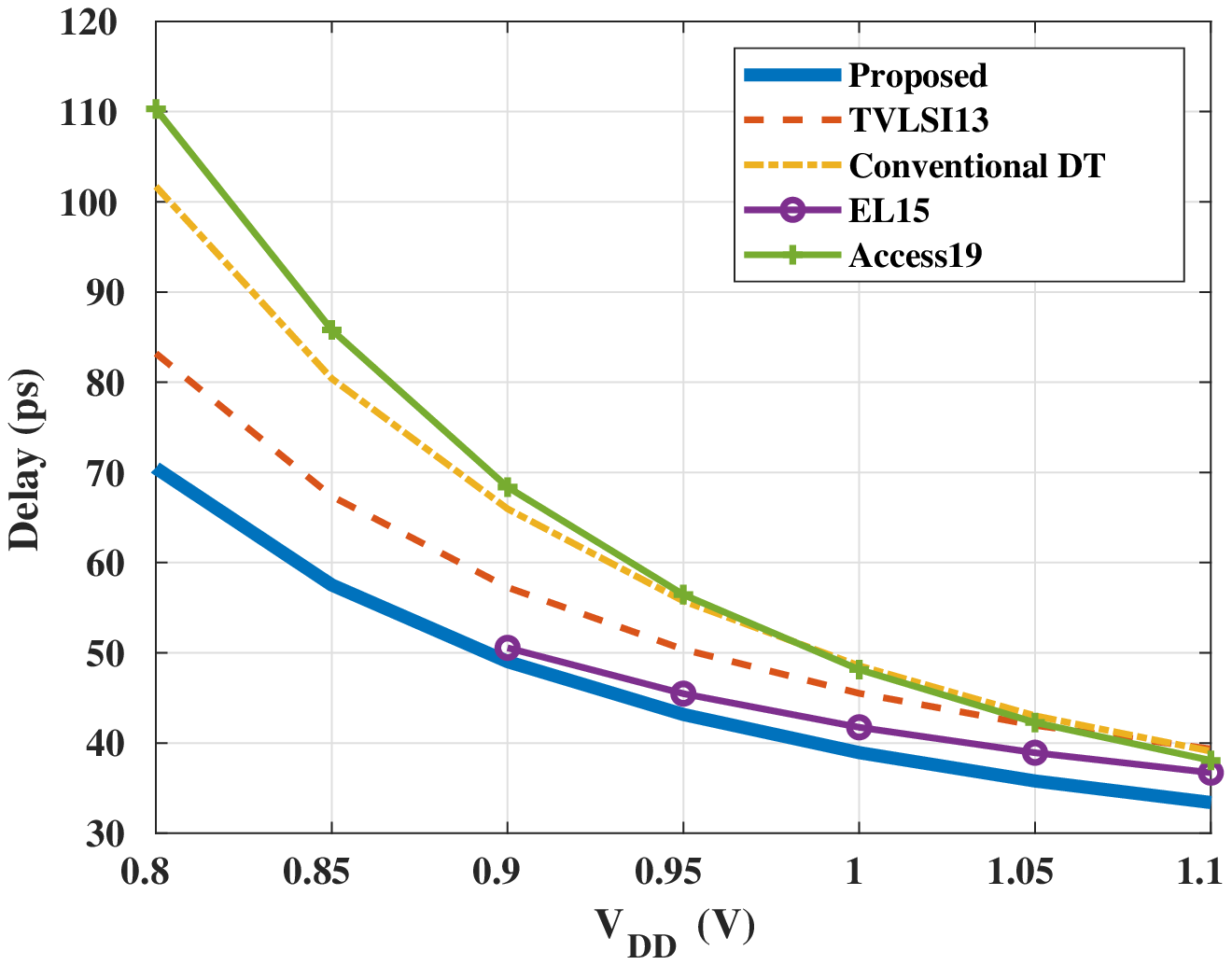}} 
    \vspace{-0.5cm}
    \caption{(a) Delay versus $\text{V}_\text{CM}$ at $\Delta V_{IN}$=10\,mV (b) Delay versus $\text{V}_\text{DD}$ of the proposed comparator in comparison to that of the state-of-the-art topologies.}  \vspace{-0.4cm}

    \label{fig_dcm}
  \end{figure}

  Fig.~\ref{fig_dcm}(a) depicts the simulated delay variation with the $\text{V}_\text{CM}$. For a sufficiently large $\text{V}_\text{CM}$, the proposed comparator is faster than the state-of-the-art comparators. Conventional DT architectures enter the triode region and thus limits pre-amplifier gain. The proposed topology overcomes this problem to some extent by achieving higher $\Delta V_O$ with the help of the cascode cross-coupled pair. Fig.~\ref{fig_dcm}(b) shows the simulation results of the delay versus the supply voltage. As expected, the delay performance worsens at lower $\text{V}_\text{DD}$. However, the plot shows that at lower $\text{V}_\text{DD}$, the speed of the proposed circuit is 30\,\% faster than the conventional one.
 
\begin{figure}[!t]
    \centering
    \includegraphics[width=0.32\textwidth]{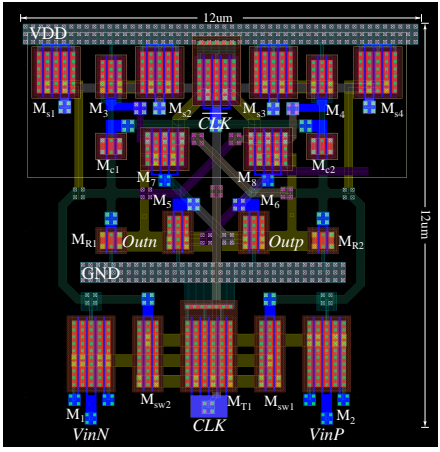}
    \caption{Layout of the proposed comparator.}
     \vspace{-0.5cm}
    \label{fig_lay}
  \end{figure}
\begin{figure}[!t]
    \centering
    \subfigure[]{\includegraphics[width=0.28\textwidth]{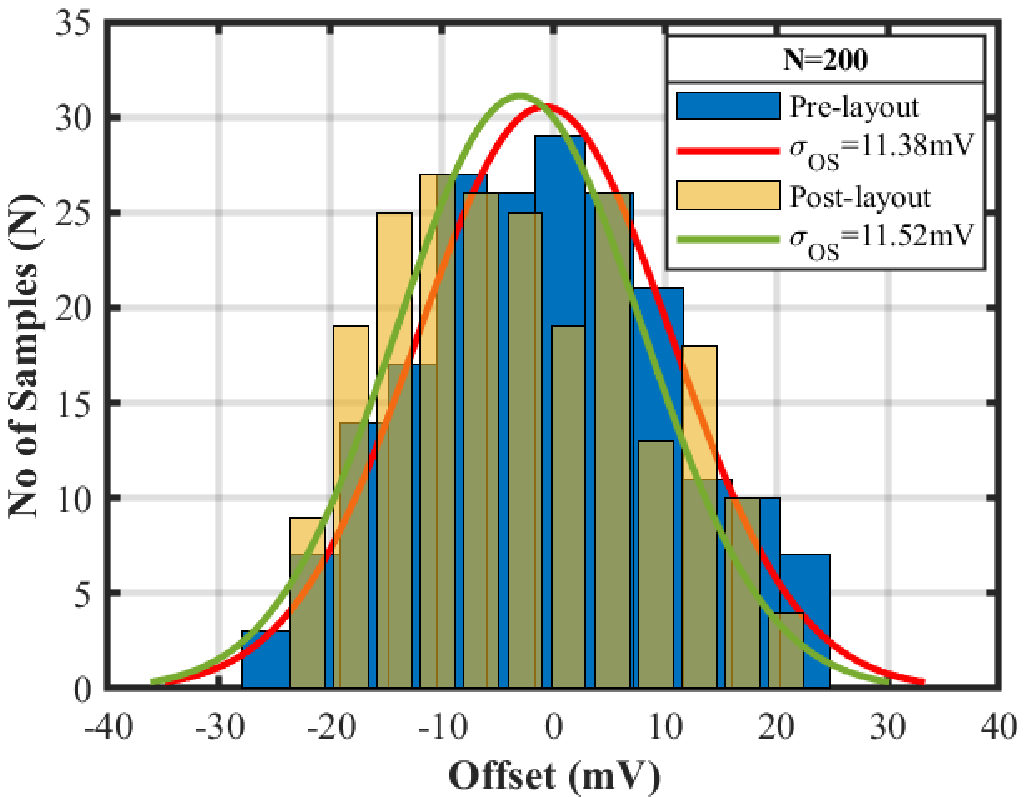}} 
    \label{MC:prelay}
    \subfigure[]{\includegraphics[scale=0.36]{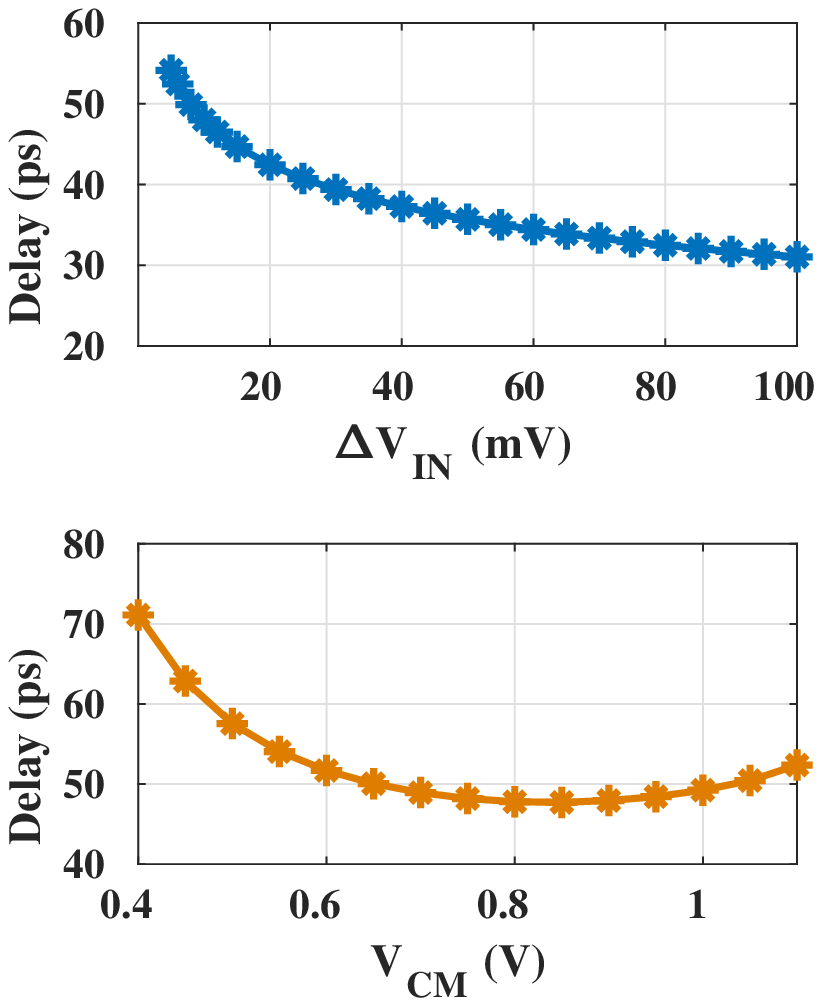}} 
    \label{MC:postlay}
    \vspace{-0.3cm}
    \caption{(a) Monte Carlo simulation results of the proposed comparator's  pre-layout and post-layout offset voltages and (b) post-layout delay versus $\Delta V_{IN}$ (top) and delay versus $\text{V}_\text{CM}$ (bottom) of the proposed comparator.}\vspace{-0.5cm}
    \label{fig_MC}
  \end{figure}

To confirm the high-speed characteristic of the proposed comparator, a layout is drawn as shown in Fig.~\ref{fig_lay} with proper care to attain symmetry with equal capacitance at the differential nodes to steer off the aspects that increase the delay of the comparator. Monte-Carlo simulations of 200 runs were performed to observe the standard deviation of the offset voltage, $\sigma_{os}$. The results are shown in Fig.~\ref{fig_MC}(a). $\sigma_{os}$ of 11.38\,mV is observed from the pre-layout simulations and 11.52\,mV is obtained from the post-layout simulations. The offset deviation is slightly more than \cite{babayan2013analysis} due to the mismatch contribution of $\text{M}_\text{c1}$ and $\text{M}_\text{c2}$.The post-layout simulation results of the delay variation with input voltage and the common-mode voltage are shown in Fig.~\ref{fig_MC}(b). An increase in delay due to the layout parasitic capacitance is evident from the post-layout results.

Various performance metrics of the proposed topology along with that of the state-of-the-art topologies are obtained for similar $\sigma_{os}$ values and the comparison is shown in Table \ref{table1}. The power-delay product (PDP) is obtained at a $\Delta V_{IN}$ of 1\,mV and tabulated in Table~\ref{table1}. PDP can be used as a figure of merit to compare the performance of the topologies. The proposed comparator has better PDP for the speed it offers. In latched comparators, the large voltage variations on the regeneration nodes cause kickback noise. It has been reported that faster comparators generate more kickback noise \cite{figueiredo2006kickback}. This trend can be seen in the proposed comparator as well. However, differential-mode KB causes no harm as long as the comparator is reset before every decision. KB also has a trade-off with offset, RMS noise, and pre-amplifier gain. Additionally, input-referred RMS noise is calculated from the transient noise simulations as explained in \cite{razavi2020design}. Input is applied and incremented in steps of 10\,uV to obtain the probability of error as 16\% to get the total input-referred RMS noise and is tabulated. 
\vspace{-0.4cm}  
  
  \vspace{-0.1cm}
  \begin{table}[!h]
    \caption{Comparison between the proposed and other comparators based on authors' simulations.}
      \label{table1}
      \centering
      \setlength{\tabcolsep}{2pt}
      \begin{tabular}[!t]{ l  c  c  c  c  c  c  c  }
        \hline\hline
        $L=60$\,\si{\nano\meter} & \bf Offset & \bf KB & \bf EC/ & \bf RMS &\multicolumn{2}{c}{{\bf Delay} (\si{\pico\second})}& \bf {PDP} \\
        $C_L=2$\,\si{\femto\farad} & $\sigma _{os}$ &\bf noise  & \bf bit & \bf noise & \multicolumn{2}{c}{$\Delta V_{IN}$} & $\Delta V_{IN}$ \\ $\text{V}_{\text{DD}}=1.1$\,V & (\si{\milli\volt}) & (\si{\milli\volt}) & (\si{\femto\joule\per bit}) &(mV) &10\,\si{\milli\volt} & 1\,\si{\milli\volt}&{1\,mV} \\ $\text{V}_{\text{CM}}=0.77$\,V&&&&&&& {(pJ.ps)}  \\
        \hline\hline
        JSSC '04\cite{wicht2004yield} & 9.9 & 2.8 & 27.4 & 1.3 & 46.8& $\dagger$& -- \\
        ISSCC '07\cite{schinkel2007double} & 10.2 & 3 & 48.1 & 1.5&39.3& $\dagger$&-- \\
        JSSC '10\cite{van201010} & 9.5 & 7.9 & 82.9 &  0.16 & 52.6&67.9& 6.9\\
        TVLSI '13\cite{babayan2013analysis} & 10.9 & 3.0 & 59.3 & 0.6 & 38.3& 59.3& 4.2 \\
        EL'15 \cite{gao2015high} & 13.6 & 3.3 & 44.9 & 2.6 & 36.7& $\dagger$&--\\
        TVLSI '18\cite{khorami2018low} & 15.1 & 51.8 & 41.7 & 3.5 & 56.7&  $\dagger$&-- \\
        JSSC '18\cite{bindra20181}$^{*}$ & 8.9 & 2.4 & 32.9 & 0.5 & 119.8& $\dagger$&-- \\
        Access '19\cite{wang2019low} & 11.5 & 3.7 & 83.3 & 0.5  & 38.1&53.9&4.98 \\
        TCASII '20\cite{siddharth20201} & 8.4 & 15.4 & 135.9 & 0.7 & 46.8& 70.4& 10.3\\
        \textbf{This Work } & \textbf{11.38} & \textbf{5.99} & \textbf{80.8} & \bf 0.75& \textbf{33.3}& \textbf{46.5} &\textbf{3.8} \\ 
        \hline\hline
        \multicolumn{8}{l}{\textsubscript{$^{*}CLK=2.5$\,GHz, $\dagger$ Comparator is not able to resolve the input.}}
      \end{tabular}
  \end{table}
\vspace{-0.7cm}
\section{Conclusion}
\label{sec:conclusion}

  In this paper, we presented a novel DT comparator topology suitable for high-speed applications. It consists of a cascode cross-coupled pair, which increases the pre-amplifier gain in the comparison phase. Furthermore, the common-mode voltage at the pre-amplifier output is lowered by the cascode cross-couple pair. As a result, the latch regenerates fast. Post-layout simulations in a 65\,nm CMOS technology with a supply of 1.1\,V confirmed that the delay is reduced considerably without much increase in the EC compared to the state-of-the-art architectures.


\bibliographystyle{ieeetr}
 \vspace{-0.5cm}
\bibliography{ref}

\end{document}